# Deterministic Dynamics and Chaos:
# Epistemology and Interdisciplinary Methodology.

Eleonora Catsigeras[1]

November 12th., 2010

## ABSTRACT

We analyze, from a theoretical viewpoint, the bidirectional interdisciplinary relation between mathematics and psychology, focused on the mathematical theory of *deterministic dynamical systems,* and in particular, on the *theory of chaos*. On one hand, there is the direct classic relation: the application of mathematics to psychology. On the other hand, we propose the converse relation which consists in the formulation of new abstract mathematical problems appearing from processes and structures under research of psychology. The bidirectional multidisciplinary relation from-to pure mathematics, largely holds with the "hard" sciences, typically physics and astronomy. But it is rather new, from the social and human sciences, towards pure mathematics.

## INTRODUCTION

The problem we focusing in this paper, is not only the application of the mathematical theory of dynamical systems to psychology, but mainly the following questions:

Which psychological processes are involved in the development of pure mathematics?

How can a multidisciplinary space be organized to activate the converse relation, from

---

[1] Instituto de Matemática Facultad de Ingeniería Universidad de la República (Uruguay).
E-mail eleonora@fing.edu.uy
Address: Av. Herrera y Reissig 565 C.P. 11300 Montevideo. Uruguay

The author was partially supported by Comisión Sectorial de Investigación Científica (Univ. de la República) and Agencia Nacional de Investigación e Innovación, of Uruguay.

psychology towards pure mathematics?.

How may psychology provide a rich field of new mathematical questions to be investigated, not only by applied mathematicians, but also by researchers on pure mathematics? Even if large advances had been achieved, the application of the mathematical theory to psychology is still mainly developed by mathematical psychologists and applied mathematicians, in the absence of pure mathematicians. Conversely, the development of the pure mathematics is now a days mainly developed in the absence of applied scientists, particularly of human and social researchers. This is the opposite situation to the antique posture, in which theoretical mathematics and philosophy, for instance, were almost a single science.

Along this paper we aim to found how the potential strength of the mathematical tools can be more fully exploited in the interdisciplinary space, and how the necessary development of new abstract and adequate tools in pure mathematics, may be detected while immersed into an interdisciplinary discussion. This discussion does not need to be "applied", in its restricted sense. In fact, mathematics may still remain abstract and theoretical, bust just break its apparent isolation from other sciences, in particular to those related with the human thinking, like philosophy and psychology.

The methodology of our analysis along this paper follows three steps: First, we present a partial review, focused in several aspects of the mathematical research, in their interdisciplinary relation with psychology. Then, we state and analyze epistemologically, the mathematical abstract definitions of dynamical systems, and in particular of deterministic chaos. Finally, we suggest a general meta-theory in the organization of the interdisciplinary space between mathematics and psychology, which we illustrate with an hypothetical example.

This paper is organized in 6 sections:. In Section 1 we briefly introduce the discourse. In Section 2 we present a partial survey of the knowledge in the interdisciplinary fields among mathematics, psychology and other sciences. That survey is focused on the theory of dynamical systems, and is very partial respect to the whole abundant development in this interdisciplinary field. In Section 3 we state the mathematical definitions of dynamical and autonomous system, and of deterministic chaos, and analyze them epistemologically. Among other properties, we revisit the argument of *self-organization* of deterministic chaos. In Sections 4 and 5 we propose a method and a meta-theory, according to which, the interdisciplinary space between mathematics and psychology may organize its purposes and actions. We consider the epistemological objection of Nowak and Vallacher (1998, p. 32). They observe that the traditional notions of causality holds in social psychological research, and oppose to (some of) the mathematical models of dynamical systems, which feedback the *same* variable from one time to the next. In §5.1.1, arguing on a particular hypothetically example, we propose a method to model mathematically such systems with causal transitions, provided that the system is deterministic. The modeling method that we propose in this meta-theory, solves the epistemological objection of Nowak and Vallacher, in some particular cases. Finally in Section 6, we state the conclusions.

## 1. MOTIVATION ▲

Ten years ago, a group of mathematicians researching on the abstract theory of deterministic dynamical systems, received an invitation to participate in the interdisciplinary discussion board of the XIV Congress of the Latin American Federation of Psychotherapy. The aim was to discuss about the theory of deterministic chaos. Therefore, their duty was to pose in exact form, but in terms that all scientists could interact, the mathematical notions about dynamics, determinism and chaos, and how these notions are conceived, created (or discovered), and evolve with the increasing knowledge of sciences. That enterprise was very motivating, an excellent opportunity to apply the pure mathematics, on which they researched, and to find new problems to study. But it really implied a challenge, a very difficult task: to understand mutually, among scientists in such different disciplines. Their procedures and purposes, were (and still are) too far. An hour of discussion in that *multi*-disciplinary board (it was not indeed an *inter*-disciplinary discussion), was not enough to shorten the distances. But since then, a rich collaborative work started to construct a necessary interdisciplinary space.

Pure mathematicians usually attribute to the object and result of their research, a platonic meaning. It is immaterial, invented by their thoughts, conceived as interior to their minds, but simultaneously felt as exterior to the creator and, most surprisingly, universally true. How so qualitatively different characteristics of the mathematical creation can be performed together? The answer is that they are not developed simultaneously. In pure mathematics the abstract definitions and deductive proofs are rigorously formalized much later to the time in which they were really conceived, created or discovered. The reader is intended to reproduce the concepts and the proofs exactly, at the ending point of the process. There is no place to discuss or to lie about the mathematical final results, after they are formalized. But, this precise and exact logic does not hold, or at least is not strict, during the previous stages of the mathematical research, while the creative processes are in advance. The mathematicians' thoughts and ideas come, before being born, from a mixture of subjective perceptions, intuitions, feelings and irrational beliefs. They are produced immersed into philosophical thoughts and influenced also by the social group in which the mathematicians act. Nevertheless, from the external viewpoint, mathematics is sometimes reduced to its final, formal and exact formulation. Summarizing, the formalism is *undoubtedly necessary* in mathematics *to state finally* the new abstract mathematical definitions and theorems and their deductive proofs. But is is clearly insufficient to perform the creative tasks of the mathematics itself, either in its pure form or in its application to other sciences. From this viewpoint, mathematics needs the knowledge from philosophy and psychology.

## 2.   PARTIAL OVERVIEW   ▲

The mathematical theory of dynamical systems is a matter of important critical discussion, about which some mathematicians reflect, particularly concerning to the philosophy of deterministic dynamics and chaos. (Massera 1988 and 1997, Stewart 1989, Ruelle 1993, Lorenz 1995, Markarian and Gambini -editors-1997, Lewowicz 2008).

On the other hand, the theory of dynamical systems is being modeling processes, for instance in social psychology:

"*Interpersonal thought and action represent highly dynamic and complex phenomena...Because of these defining qualities, social psychology is highly amenable to understanding and investigation within the framework of dynamical systems theory.*" (Vallacher and Nowak 1997).

In the same article the authors raised the relevance of a *meta-theory:* how to apply the theory of dynamical systems in the research of psychology, mapping the abstract notions onto established psychological phenomena and models. In this paper, we focus on that viewpoint, and also in the converse one: how the established psychological dynamical phenomena may inspire new abstract problems to be investigated by mathematics.

We propose a method to support the interdisciplinary space and to bridge the research actions between the two disciplines. In the same line Scott (1994), discussed the relations between the terms defined in the models of psychological dynamics and the concepts of chaos and self-organization. Nevertheless, he recognized the difficulties in using the abstract notions and methodologies of mathematics in empirical investigations of psychology. On the other hand Eliasmith (1996) presented a critical examination of the dynamicist theories in some fields of psychology, in particular to the area of cognition. Also Ayers (1997) examined the applications of chaos in the research of cognitive developmental and clinical psychology and their possible implications, evaluating problems regarding the usefulness of chaos in psychology. Robertson (1995) established a wider spectrum of applications of the theory of chaos to psychology and life sciences.

To understand and explain how the brain performs processes (for instance learning, memorizing, and associating sensory manifestations with memory) the structure and dynamics of the neural system and its mathematical models play a fundamental role. The precursory paper of Kohonen (1977). examined the associative memory from neurophysiology and psychology. Later, the related mathematical models were intensely studied (see for instance Cooper 1995, cited in Mizraji 2010). More recently, those models and some of its consequent generalizations derived in new relevant conclusions on the associative memory process, its applications to psychology and its philosophical implications (see for instance Mizraji 2007 and 2008, and Lansner 2009).

From the advances of the mathematical models of neural networks, the study of neuro-dynamics has given a new insight and development. The first mathematical model of neurons was given by the differential equations of Hodgkin and Huxley (1952, cited in Lamberti and Rodríguez 2007). After that, almost four decades passed until the first rigorous mathematical proofs appeared, using the abstract methods of the theory of dynamical systems. They derived into new theoretical results on the dynamics of neural networks (Mirollo and Strogatz 1990, Budelli et al. 1991 and 1997, Budelli and Catsigeras 1992, Rieke et al. 1997, Coombes and Lord 1997). Later the general models of general impulsive controlled coupled oscillators raised, including the mathematical dynamics of neural systems as particular cases. (Timme et al. 2002, Ishikevich 2007). Also the algebraic and topological self-organization of neural networks was well described in many complex cases. (Mizraji 2007, Coombes 2007, Coombes and Laing 2009).

In the last decades many mathematical results about the dynamics of some particular neural subsystems found only periodic and stable behaviors, and so the temporal variation seemed to be simple. Nevertheless, some computer simulated and deductively proved results about non periodic, irregular and chaotic dynamics in neural networks appeared (Feudel et al 2000, Timme et al. 2002, Catsigeras 2010). An apparent contradiction raises from the fact that some models of neural systems are non chaotic, stable and periodic, according to the classic mathematical proofs, but they appear irregular in computational experiments. This apparent contradiction was explained by Cessac (2008): the theoretical periods may be extremely large, out of the scale of time of the experimentation or of the life of the system. So, even if they are not strictly chaotic according to the mathematical definition of chaos (which requires infinite time), the experimenter can observe only the irregular transitory behavior. This is called a *virtual chaos*.

The knowledge of the neuro-dynamics allowed also the construction of systems of artificial neural networks. These last, as well as the biological neural networks, opened for mathematics a new spectra of problems to be solved. They deal, for instance, with chaos, or unstability, of systems that usually exhibit discontinuities and very large dimension.

The dynamical theoretical results obtained from the mathematical theory, are finally decoded, to explain some deterministic psychological processes. Besides, the man-created neural networks are partially inspired in the physiology of biological neural systems. Some of them try to reproduce deterministic psychological processes, such as learning and memory. They are widely applied to control engineering, to the design and investigation of artificial intelligence, and to modern communication systems (for instance in Yang and Chua 1997).

The algebraic theory of the mathematical *laws of the thoughts* by George Boole (1854, cited in Camacho 2006 and Mizraji 2010), based the modern development of digital systems in informatics and electronic engineering. The mathematical relations between *the computer and the brain* posed by John von Neumann (1958 posthumous), started the interdisciplinary interaction between mathematics, engineering and psychology. This interaction is increasing in relevance, and besides, has amazing applications to informatics and technological disciplines:

*"Different cognitive networks are built up by different human individuals, all of them sharing the same large scale brain circuitry. It is interesting to mention here that this fact suggests a metaphorical analogy between natural neuronal networks and some technological information networks. Thus, we can put in correspondence the neuroanatomy with the Internet on the one hand, and the brain cognitive network with the World Wide Web (WWW) by the other....we show how, perhaps unexpectedly, this analogy is accompanied by structurally similar mathematical models concerning information retrieval in neural memories and in the WWW."* (Mizraji 2008).

The example above is paradigmatic of the closing interdisciplinary loop among many disciplines, including mathematics and psychology. Nevertheless we note that the human psychology is far from being fully deterministic. In fact, this question revitalizes the old philosophical paradox of determinism:

*"...we demonstrate that, even in a deterministic universe, there are fundamental, non-epistemic limitations on the ability of one subsystem ... to predict the future behaviour ...in the same universe. ...These limitations arise because the predictions themselves are ...part of the law-like causal chain of events in the deterministic universe. ...Even in a deterministic universe, human agents have a take-it-or-leave-it control over revealed predictions of their future behaviour."* (Rummens and Cuypers 2010).

An old mathematical conjecture states that uniform global hyperbolic diffeomorphisms on compact manifolds (which are the diffeomorphic paradigms of deterministic chaos), can only evolve under very strong topological restrictions of the space. Related with this open problem, twenty years ago Jorge Lewowicz proved in dimension two, that the expansive systems (which are all the topological chaotic systems) can only exist under some non trivial restrictions of the topology of the space where they evolve (<u>Lewowicz, 1990</u>). His proofs suggest how can be extended to larger finite dimensions. Later they have been generalized to dimension three (<u>Vieitez, 1996</u>). Nevertheless for larger dimensional expansive systems the conjecture is open. On the other hand, for very large scale physical systems it is unknown if those topological restrictions hold. With more reasons, it is unknown if the global system of individual or social human psychology, or its subsystems, satisfy them. So science, not only psychology, but also theoretical physics and pure mathematics, can not still invoke the deterministic chaos to explain all the apparently irregular or unpredictable dynamics.

The old paradox of determinism and predictability, versus chaos and unpredictability, also appears in the following text, written in 1864, almost a century before the mathematical theory of deterministic chaos raised into the discourse of the hard sciences:

*"If you say that all that can be predicted: the chaos, ... that the mere possibility of a previous calculus can contain everything, and that the* (rational determinism) *will end to prevail, then the man would become mad on purpose, for not to have the truth and behave according with his wish."*
(Fyodor Dostoyevsky, Notes from Underground, 1864, cited in <u>Lewowicz 2002, p.53</u>).

### 3. MATHEMATICAL DEFINITIONS OF DETERMINISM AND CHAOS ▲

Unfortunately, when mathematicians and physicists gave a name to the particular *strange* phenomenon appearing in some dynamical systems, they called it *chaos*, or *deterministic chaos*. But, as we will explain below, the mathematical deterministic chaos does not fit with the usual meaning of the name *chaos*. In fact, the mathematical deterministic chaos is indeed well ordered, with zero degree of confusion, no fuzzy behavior, and is rather well understood now a days. It is non hazardous. It is governed by defined rules. In brief, a mathematical system exhibiting deterministic chaos is *self-organized*. The name chaos is due only to the fact that an observer of the dynamical phenomena, may not perceive a priori its organization. For instance, who sees on the monitor of a computer the orbits by successive iterations of the one-dimensional quadratic law

$$x_n = 4\,x_{n-1}\left(1 - x_{n-1}\right) = 4\,x_{n-1} - 4\,x_{n-1}^2 \qquad (1)$$

will perceive them as disordered. They appear chaotically, in the usual sense of this word. So, the observer sees them, a priori, as unpredictable and hazardous, in the same way that a person, who does not understand the Chinese language, could perceive the sounds of someone talking in Chinese, as disordered and unpredictable.

If the precise definitions and hypothesis of work are not understood, the later blind application of the results of the theory to other sciences may lead to mistakes and to ungrounded inductions (Ruelle 1990, Goldstein 1995, Sokal and Bricmont 1999, Kellert 2008). Mathematics, and in particular the theory of deterministic chaos, is not an exception from what the literal meaning of semantics implies:

*"Sometimes the meaning of a sentence is such that its truth conditions will vary systematically with the contexts of its literal utterance."* (Searle 1978).

Thus, it is necessary to precise the truth conditions, in our case the mathematical definitions, to fix the context in which the mathematical theorems about deterministic chaos are true.

**Definition 3.1. Dynamical system**

A dynamical system is a mathematical structure that admits many potential *states*, and each state is described by numerical or non-numerical *variables* which change (namely *evolve*) with time, according to the following *deterministic hypothesis*:

**Definition 3.2. The deterministic hypothesis**

The deterministic hypothesis of a dynamical system assumes *the existence of a law L* (even if it is unknown) that is not hazardous and *governs the evolution with time* of the *variables* of the dynamical system such that:

•       The previous state of the system is *the unique cause that determines* the next state of the system, while times goes on.

•       The *law L is* the exact *mathematical rule or set of rules* that *transforms* the previous state onto the next one.

In a deterministic dynamical system, the same previous cause (or set of causes) produces the same posterior effect (or set of effects). The exact law $L$ is the abstract rule or set of rules that transforms the cause onto its correspondent consequence. The cause is the state $x_{n-1}$ at the previous instant denoted $n\text{-}1$. The effect is the state $x_n$ at the next instant $n$.

Thus, after applying the law *L* to $x_{n-1}$, the obtained result is $x_n$. This is denoted as:

$$L(x_{n-1}) = x_n$$

Inductively, the state $x_n$ will be later the cause that will produce its next effect, namely the state $x_{n+1}$ in the future instant $n+1$, Precisely: $L(x_n) = x_{n+1}$.

Therefore, the following of the following state is the result of iterating the law two times. And the following of the following of the following state will be the result of iterating the law three times. This process can be repeated finitely many times, as many times as wanted.

Consequently, the state at instant $n$ is the result of iterating the law exactly $n$ times. After applied to the initial state $x_0$, the state at instant $n$ is:

$$x_n = L\big(L\big(L...\big(L(x_0)\big)\big)\big) = L^n(x_0).$$

In the notation above the exponent $n$ means the iteration of the law $\boldsymbol{L}$ repeated consecutively $n$ times.

For instance, a simple case is that in which the space of states is the set of real numbers between 0 and 1. In this case the law $\boldsymbol{L}$ can be, for example, that one defined by the equality (1) at the beginning of this section. This law, which is called *quadratic*, admits a formulation in terms of the numerical variables. So, it can be represented as a curve graph. In the example of the quadratic law, this graph has the form of an inverted U (a parabola) in the cartesian plane. But this is just an example and extremely simplified. In fact, *most mathematical laws or transformations defining dynamical systems, do not admit a formulation in terms of a curve or graph in the plane.* The variables and the deterministic law governing them, need to be defined and studied in non numerical abstract structures such as, for example, functional spaces, measure-spaces, abstract algebras, general topological spaces or geometric general manifolds.

**Definition 3.3. Autonomous system**

Under the deterministic hypothesis, the dynamical system is called *autonomous*, if the law $\boldsymbol{L}$ is invariant with time. Thus, $\boldsymbol{L}$ remains *unchanged* while time, and therefore the state of the system, may change.

The example given by the equation (1) is an autonomous dynamical system: in fact, the equality (1) itself remains the same. The *specification of the operations* that affect the value of the variable $x_{n-1}$ to lead to its next value $x_n$, is always the same, regardless that the values of $x_{n-1}$ and of $x_n$ may change with the time $n$.

The *trajectory or orbit* for each initial state is defined as the sequence of consecutive states $x_n$ of the system starting in the initial state $x_0$. This definition holds for the so called "discrete-time dynamical systems", since the instants ..., $n-1$, $n$, $n+1$, ... are computed only with values in the set of integer numbers. Therefore, the trajectories are not necessarily curves, but *sequences of* *values of the variables*, and these values are not necessarily numbers, but may be other kind of mathematical, *non numerically defined objects.* Even in the most abstract cases, the  states or values of the variables are called *points. B*ut usually, they are not  literally points, in any geometrical sense nor values in any numerical sense.

The trajectory is a sequence of points.  *The set of all potential states or points is called space.* So, the trajectories or orbits are sequences of points in the space, although this space is an abstract object that has not necessarily a geometrical or a numerical description. The trajectory is said to be *in the future* when the instants *n* are chosen larger or equal than zero. But if the states of the system were also defined for negative instants, then the trajectory *in the past* is similarly defined.

**Definition 3.4. Deterministic Chaos**

A deterministic dynamical system is said to be  *expansive, or sensitive to initial conditions or chaotic,* if two different initial states, regardless that they may be arbitrarily next, define trajectories that separate one from the other, at some future or past time, more than a positive numerical constant $a$ .

To the observer,  the positive number  $a$  is the  *perceptible*  threshold of the error caused when taking one different initial state instead of other, regardless of how next the two different initial states are one from the other.

We notice that, even if the different (potential or effectively reached) states of the system, do not need, in the modern mathematics, to be described by numerical values, the threshold of the error is indeed described by a positive (large or small) real number $a$.  For that purpose, the space of all the states, which is not numerically defined, is provided with a metric structure. This metrizability of the space allows the observer to compute the distance between two different states, even if those states may be non numerical, quantitative described, and very complex mathematical objects.

The deterministic chaos implies the uniqueness and distinguishable evolution of each individual trajectory in the system. That is why it is *unpredictable,* i.e. a single trajectory can not be *completely* predictable for all future or past times, unless all the initial data of that individual trajectory is exactly known. But, if the number of individual trajectories is too large, or infinite,  the probability to know *exactly* the initial data of one of them, is usually equal to zero.

For example, the system given by the quadratic equation (1) is chaotic. Two different initial states can be taken such that their numerical difference is arbitrarily small. But nevertheless, for some instant *n,* the difference of the two states, evolving along their respective trajectories, according to the quadratic equation (1), will take values as near 1 as wanted. This assertion can be proved rigorously by the mathematical theory, using the deductive method founded in the classic logic, without making experiments in a computer. The proof is far from being immediate or easy. This simple example of the quadratic law is numerical. But also in many complex systems that are non numerical, the existence of chaos can be proved using the deductive method of pure and classic mathematics. Nevertheless, this task is usually very difficult. The existence or the absence of chaos, in most such examples, and even if some of them have a very simple statement of the law $L$, is still unknown. In fact, in most complex examples, this problem is a mathematical question to which no human being has still discovered or invented a *rigorously proved* answer: *yes* or *no*. This mathematical problem has nothing to do with the unpredictability of the system under study. In the classic logic, which rules mathematics, the system is either predictable (non chaotic) or unpredictable (chaotic). Even when nobody knows the answer for a particular system, if it is chaotic or not, for the mathematics, this exact answer does exist, and is either yer or no. But the determinisitic unpredictability, of the system itself, is a different question. If the system is unpredictable or chaotic, then it is *certainly* true the following assertion:

If the initial condition *is not exactly* $x_0$ , then in some time $n$ in the future (or in the past) the state $x_n$ of the system will *certainly* differ from the expected one, more than the *perceptible* error $a$.

So, even if it seems a contradiction, the concept of unpredictability in the mathematical theory of the deterministic chaos, is indeed a certain prediction: it asserts that the observer will (not probably, but surely) make a perceptible error, if he tries to predict the evolution of the chaotic system, for which each individual trajectory is different from all the others, without knowing exactly all the initial data of such trajectory.

But not all the hopes are lost when a mathematician investigates a chaotic system. In spite of the unpredictability of chaotic systems, the so called *ergodic theory* proves that, under rather general additional hypothesis, there exists a decomposition of the space into abstract *invariant-measure structures,* called *ergodic measures.* (Mañé, 1983, pp. 162—172). They are spatial and theoretical measure-structures such that, under the optic of each of them, the states of the system evolving with time are statistically predictable at infinite time. Thus, after one of the ergodic measure-structures is chosen, the newly defined unpredictability of deterministic chaos, disappears. Statistically ergodic predictability of deterministic chaos raises due to the following two new viewpoints:

- On one hand, the observer is aimed to predict the evolution of a significant set of trajectories, instead to predict only one, among non countable infinitely many. Those significant sets may be, anyway, arbitrarily small.

- On the other hand, the way of measuring the sets is adapted to the dynamical behavior of the system. This spacial measure-adaptation is given by one of the ergodic measure-structures. Nevertheless, the ergodic measure-structures of the space disregard the transitory states, which do not lay on the supports of the ergodic measures, i.e. on the attractors. The ergodic measures consider only the so called regime states, which are the asymptotic states in the future, supported on the attractors.

We conclude that the *mathematical deterministic chaos* is not literally chaos, and its name is just wrongly chosen. Some more adequate names, that are not so popular, but that are certainly used by pure mathematicians now a days, are *expansitivity* if referring to topological chaos, *mixing* if referring to a strong form of topological or measurable chaos, and *hyperbolicity or existence of positive Lyapunov exponents*, if referring to differentiable chaos, in which there is an exponential rate of expansion along some subspace.

One of the most notable examples of deterministic systems related with psychological processes is the dynamics of some models of neural networks. Other deterministically modeled psychological process, is for instance the deductive rational thinking, which is formalized by the boolean algebraic rules. That is why it is possible to reproduce it through computers and artificial intelligence. Pattern recognition is also a clear example of application to psychology of the mathematical theory of dynamical systems. This application has developed tools to explain how patterns can be identified in what otherwise could be interpreted as an hazardous temporal variation. Other example of how the mathematical theory can model a psychological manifestation, is the quadratic law (1). Performance effectiveness may depend on the values of the variables in the form of an inverted U function: "*it increases up to a point beyond which further increases in arousal promote performance decrements*" (Nowak and Vallacher, 1998, page 36) Such a relationship is posed by Atkinson's law, in a model of some type of simple motivation following the quadratic function. The motivation increases with the difficulty of achievement the goal, up to a certain level of difficulty. After that level, it decreases because it becomes inversely dependent of the probability of success. Therefore, the motivation function takes the form of an inverted U, which produces a maximal value when the difficulty of achievement takes half of its maximum value.

Nevertheless, many psychological systems do not satisfy the deterministic hypothesis. According to this hypothesis, the same variable (the state of the system) "*acts as a cause one* *moment and as an effect the next*". In fact, the sequential iteration of the deterministic law can be interpreted as the feedback of the *same* variable from one instant to the next. "*This feedback process is at odds with traditional notions of causality that assume asymmetrical one-directional relationships between cause and effects. For the same reason, it does not fit well with* (some models of) *social psychological research.*" (Nowak and Vallacher, 1998, page 32). To solve this problem, we propose to use a model of dynamical system evolving on a more abstract mathematical structure, such as a *functional space*. In that space the variable (the state) does not represent directly the psychological manifestations. These manifestations depend of intrinsic temporal conditions and of external causal agents. It is a response of a complex combination of those conditions or agents, among which the previous manifestations are not the unique ingredient. In a mathematical functional space, the variable to be analyzed is no longer the response, the manifestation, nor the agents that cause them, but *the way* in which the agents cause the response. Therefore, the feedback of this functional variable one time, into the *same* functional variable the next time, takes into account, embedded in the functional structure, all the factors varying with time. The feedback iterative process does not restrict its incomes to the response manifested at the previous time. In §5.1 we expose an illustrative example of this abstract modeling, in the case of a social psychological process.

## 4.    GROUNDING AN INTERDISCIPLINARY SPACE    ▲

From the viewpoint of the classic logic of pure mathematics, the following schema holds:

### 4.1 In pure mathematics

A mathematical theorem $(T)$ states

$$(T): \ A \Rightarrow B$$

The symbol $\Rightarrow$ denotes the word "implies". The assertion $(T)$ means that there is a mathematical proof, which is published and exhibits by rigorous deduction, that the hypothesis $A$ implies the thesis $B$. But, when occurs $B$, nothing can be said, if *only* theorem $(T)$ is invoked, about the occurrence of $A$. Theorem $(T)$ does not explain why $B$ may occur in all the cases. In other words, even if $B$ could be exactly the same phenomenon, appearing for instance in *all* the observed psychological process of some type, and reported as a thesis of a mathematical theorem $(T)$, this theorem *does not explain* the appearance of the phenomenon $B$ in all the known or unknown cases, unless the hypothesis $A$ holded surely for that type of processes.

### 4.2 The First Direct Task

The first task when applying a mathematical theorem $(T)$, is to check (or to assume) that the epistemological characteristics of the particular system under study (for instance the dynamical system governing a psychological process), satisfies all the assertions *included in the hypothesis $A$.* In other words, for instance if $A$ is a deterministic system modeling the psychological process, it must model this process without loosing its characteristics, being a representative of the *same system* under investigation of psychology. The achievement, when applying Theorem $(T)$, is not only to fit the thesis $B$ a posteriori, *but fit the hypothesis $A$ a priori.* And besides, a mathematical theorem can be applied to the system under investigation by psychology, if its hypothesis $A$ represents the system under study, *without oversimplifying it.*

### 4.3. In psychology and other sciences

Let $\widehat{A}$ denote a social or individual psychological dynamical system under study, which exhibits, after observation or psychological theoretical research, the features identified as the phenomenon $\widehat{B}$. We denote this experimental or theoretical result of psychology research:

$$(R): \widehat{A} \rightarrow \rightarrow \widehat{B}$$

The symbol " $\rightarrow \rightarrow$ " denotes that the scientifically grounded action to pass from $\widehat{A}$ to $\widehat{B}$ after an observation or theoretical research in psychology. All the known psychological systems of the class $\widehat{A}$ exhibit the property or behavior $\widehat{B}$. In other words, after all the observations of psychology, was never found $\widehat{A}$ and *not $\widehat{B}$*. Thus, it is scientifically *induced ,* in psychology and also in other sciences, that $\widehat{B}$ occurs if $\widehat{A}$ occurs. In other words, $\widehat{A}$ causes the phenomenon $\widehat{B}$ in all the cases, because the universe of all the cases is, in the human and social sciences, the collection of all the  *known* cases.

  The inductive method as described here, is undoubtedly legitime and scientifically valid, not only because it is widely used in most sciences (but not in mathematics), but because it gives strictly objective evidence, to increase the human knowledge about the general nature of the system under study. But it is not classically accepted to constitute a rigorous mathematical proof, which must be strictly deductive according to the rules of the classic logic, and do not have recourse on the induction method.

### 4.4 The Second Converse Task

   At one side, the known mathematical results in the theory of deterministic dynamical systems, are still few and narrowed, to *explain* many relevant problems of other sciences, in particular of psychology. At a second side, but not less important, many scientific objects of research of other sciences seem that can be mathematically modeled as deterministic and chaotic phenomena. Thus, joining the two sided aspects, it raises  the following *challenge to applied and pure mathematics*:

•      (i) To state the new problems that arise from other sciences, which are mathematically translated or modeled, with a precise formulation.

•      (ii) To investigate them,  under the mathematical classical methods, that is, without giving up to the deductive mathematical proofs founded in the classic logic, but also without making the other sciences to give up of their own methods. Schematically, the mathematical investigation searches for:

•      (ii)-1. Definitions of the mathematical and abstract concepts that fit with the structure of the applied problem.

•      (ii)-2. Hypothesis that do not oversimplify or restrict the applied objects under study to a set of empty relevance in the other science.

•      (ii)-3. Mathematically posed conjectures, including those relations obtained by induction from the results of the other science.

•      (ii)-4. Theorems: conjectures, after proved to be true, by means of a rigorous deductive method based in the classical logic.

•      (ii)-5. Mathematical counterexamples, which prove that a conjecture is false. The mathematical counterexample must not necessarily represent a real observed example of the other science.

•      (ii)-6. Reformulations of  the mathematical results   to be   applied to the problem under

study in the other science, and explain or predict them.

Mathematicians do not expect to perform the six activities of the list above, ordered in time, in the same numerated sequence of the items (ii)-1 to (ii)-6. During the research work, the attempts to prove a conjecture, may derive in the revision of the definitions of the concepts and hypothesis, which are only provisional until the work is ended. Sometimes, new attempts of proofs, derive in changing again the purposes, the conjecture itself, which may derive in a new reformulation of the abstract object under investigation, and of the strategy of research.

The third-exclusion principle in mathematics establishes that *each mathematical assertion is either true or false*. Thus, a mathematical *conjecture* (C), is an *open question:* no mathematician has already prove it nor refute it. The refutation consists in finding and exhibiting a counterexample. If no counterexample and no proof have been discovered and published then the conjecture (C) itself is not a new a mathematical result. It is not new, even if hypothetically a mathematician discovered a billion of examples for which (C) is true and that his billion of examples were *all the known examples.* On the other hand, in psychology and most other sciences, the third-exclusion principle and its derived practical rules do not hold. Hypothetically a psychologist who could show that all the human beings in this moment living on the Earth behave according to the assertion (C) would have a new result of extremely large relevance.

If the object of research in mathematics is not coming from a problem posed by other sciences, a change of strategy is almost always used, which is undoubtedly legitimate: *to make stronger or more restrictive the hypothesis* of research, to simplify the mathematical problem under investigation as much as needed (and as less as possible). In this way the hypothesis are weakened to fit to the new deductive proofs that the mathematician could find. So, he can obtain a new mathematical result. Therefore, the mathematician proves a *weaker theorem*, instead of other more difficult and general, whose proof or counterexample is assumed to exist, but remains open.

We are not referring here to that legitimate strategy (fit the hypothesis to a new known proof), when we are trying to pose the difficult reciprocal task in applied mathematics. When mathematicians research some of the problems derived from other sciences, in particular from psychology *the challenge* is the following:

Applied mathematics research develops in such a way that the *hypothesis fits with the problem* to which the mathematical results are going to be applied. Thus, the strategy of taking stronger hypothesis, even if always legitimate and very useful for the advances in pure mathematics, is not always useful for the further applications of the mathematical result so obtained.

## 5. AN INTERDISCIPLINARY METATHEORY ▲

As argued in the last section, the purposes and scientific methods of psychology and mathematics, differ ones from the others, since their own epistemological basis, but do not need to give up to those own differences, to be able to interact. The philosophical diversity in sciences (Zollman 2010) analyzes the epistemology of inter or intra-discipline, when there are different scientists working in the same problem, but with different approaches and methods. Without taking a position about what are called *unified or diversified scientific methods*, we will pose a third metatheory:

Instead of the extremes of homogenizing the science, and of atomizing it into many almost disconnected disciplines, a collaborative interdisciplinary interaction can be developed. Its methods and purposes have to be collaboratively defined, and may differ case by case. The interdisciplinary space is neither the simple sum of the purposes and methods of different sciences, nor the over sized prevalence of one science over the others, nor the joint homogenization of many science without differential purposes and methods. The multidirectional connectivity among them is established in such a way that none of the different sciences had to give up to its particular epistemological identity. (See Figure 1)

### 5.1 Modeling the structure and dynamics of a psychological system

Revisiting the notations used in the paragraph 4.3, let us denote $\widehat{A}$ to the complete set of structural characteristics of a certain *class* of hypothetical (individual or social) dynamical system under investigation of psychology.

One of the actions in an interdisciplinary space between psychology and mathematics, is to *translate* the structural characteristics of the class $\widehat{A}$ of systems (to fix ideas assume that those systems are all observed psychological and social types of groups of workers, for instance), into a mathematical model $A$. This action considers abstractions of the qualitative manifestations of the system. When we refer to a *mathematical model* we are not restricting mathematics to calculus, to numerical methods, to numerical computation, to statistics, nor to the theory of dynamical systems evolving in a finite dimensional numerical space or geometric manifold.

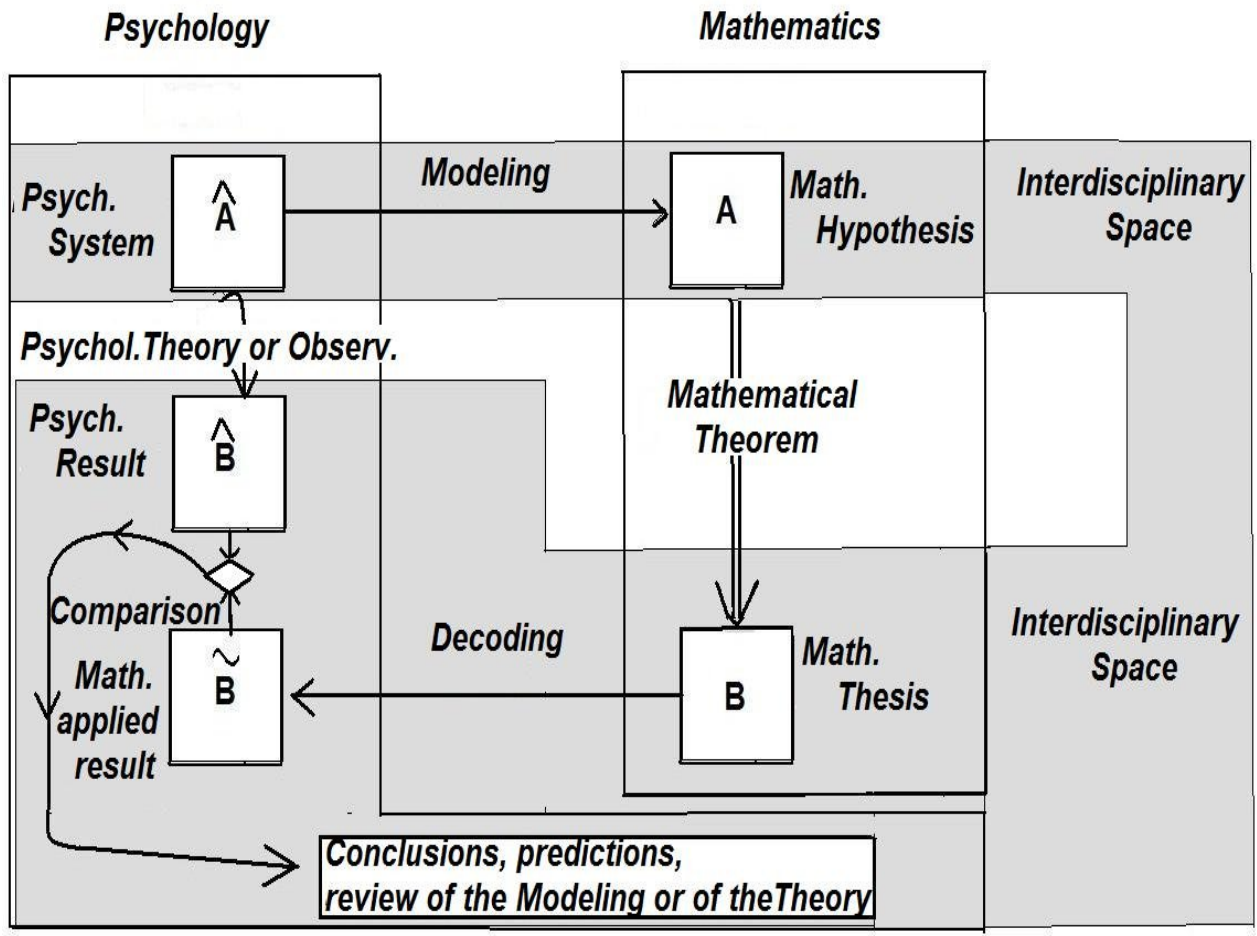

Fig. 1 Blocks schema of the inter-disciplinary space.

The *mathematical modeling* of the psychological system will be denoted as

$$\widehat{A} \to A$$

The modeling action is denoted with the single arrow $\to$, which is one-directional. In fact, it is enough that all the characteristics of the psychological system $\widehat{A}$ are translated into its mathematical model, the system $A$, but the converse relationship is not required. Indeed, if our purpose is to study $\widehat{A}$, then its mathematical model $A$ acts just a tool. As any tool, *all* the specifications of $\widehat{A}$ may be considered, but the tool may have also some other characteristics that are independent of those of the object to which it serves. Namely, some extra conditions of $A$, neither represent the characteristics of $\widehat{A}$ nor contradict them.

To fix ideas, let's put an example by analogy:

A hammer is a tool to fix nails. The hammer has a a head whose characteristics must fit well with *all* the nails to be fixed. But it also has a handle, which may have a form to fit with the worker's hand. The handle's form fits with the fingers of the worker, but are meaningless with respect to the nails. Nevertheless, the handle must be *compatible and not restrictive* with the action of fixing *all* the nails. For instance it can not be made with a too soft material, which would fit nicely to the worker's hand, but would bend with the use and make the hammer useless for its purpose.

To end the analogous idea, we mean that sometimes, the mathematical model would be nice to obtain interesting theorems for pure mathematics, but inadequate or oversimplified, if the variables, parameters and constants do not fit well with *the complex characteristics* of the system under study, to which the thesis of the theorems are supposed to apply. In particular, the most common oversimplification that may make useless the theoretical results, is that of modeling mathematically the complex systems under research of other sciences, using only real numbers or geometric approaches in finite dimensional spaces. This oversimplification is particularly abundant in the bibliography about the mathematical models of systems coming from psychology.

### 5.1.1. An hypothetical example. ▲

Consider the hypothetical translation into a mathematical structure of the psychological manifestations of a social group, describing their attitudes and behaviors respect to work or job activities. One needs to consider many complex and *qualitative* descriptions: competitions and pressures, individual and collective wishes and expectations, social levels, motivations, scholar education, qualification for the work, abilities, external opportunities and the psychological perceptions of them, achievements, goals, etc. Also the relations among all those variables had to be considered as variables. Besides the tendencies of change of each variable may be considered also as part of the variables, and the agents that may influence in those tendencies, too. Some of the variables are intra-individual, other are extra-individual but intra-social-group, and other are extra-social-group. Let us consider a mathematical model with a complexity comparable with that of the system itself. Our proposition is hypothetical, and just intended to be illustrative. Besides, it is not the unique possible model.

First, recall that, in general, the mathematical variables are neither numerical, nor a finite set of real numbers. Most variables considered in the modern mathematics, are so abstract that do not live in numerical fields or geometrical spaces. Unfortunately, the mathematical theory of chaos and of dynamical systems, is still almost completely undeveloped in such general abstract spaces.

Both disciplines, mathematics and psychology, require and use a high level of abstraction. So, let us take advantage of a sequence of increasing levels of abstraction: Consider an abstract *transformation* $X$ such that, to each codified entrance gives some defined codified output. Each entrance, and also each output, is not necessarily codified with numbers, and there may be infinitely many possible entrances and/or outputs. The set of all the codified potential entrances is called the *domain* of the transformation $X$, and the set of all the codified potential outputs is called its *co-domain.* The transformation $X$ itself is also called in mathematics *a function*, even if its domain and co-domain are not sets of real numbers, and so, its graphic curve or line, as a set in a cartesian plane, *is not defined.* Most mathematical functions *are not* identifiable with such a graphic in the plane nor in a finite dimensional space. Abstract functions, as defined above, are mathematical objects as well as those that have numerical formulations.

Uploading one more abstract step, such transformation $X$ does not need to be unique and remain static, but may be a variable state, varying in a set of many *potential* states of a dynamical system. So, each function is a "value" of the variable $X$. Therefore, this variable $X$ takes *values* in the set of all the transformations described as above. We denote this set as $\aleph$ (aleph), and is said to be a *functional space*.

Returning to the hypothetical example of psychology, the set of all possible attitudes of the social group is the co-domain of the function $X$ modeling the attitudes' reactivity function at one time. Therefore, $X$ does not model the attitudes separately, nor the agents that cause them, but the relations among the agents and the attitudes. The space $\aleph$ is the set of all the potential attitude's reactivity functions of the group, as can be conceived by the psychology researcher, regardless if they will or will not effectively appear.

Let us upload one more step of abstraction. Consider the deterministic dynamical properties of the system, according to Definition 4.1. The instantaneous state is $X$ and is also called *point,* being in our example an abstract function, that is the attitudes' reactivity function of the social group on one fixed time. Embedded in the variable $X$, which is called "one point" in the functional space $\aleph$, are all the agents, intra or extra-individual, and intra or extra-social group, all the qualitative (not only the quantitative) responses of the social group to all those agents, and only one functional way in which all the agents cause all the responses. Thus, $X$ is the reaction (or the way to rspond) of the social group, at one instant $n$, and changes, or "moves" with time, inside the functional space $\aleph$. It is imagined as a point-wise fly in the air, disregarding that its mathematical definition is much more complex than that. The point $X$ is changing, as time goes on, moving or evolving in the space $\aleph$ of all the possible states. As defined in 3.2, if the system is deterministic, there is a dynamical law $L$, which is also a transformation, whose domain is now $\aleph$ and its co-domain is also $\aleph$. The dynamical law $L$ transforms the point $X$ at one instant $n$ onto the following point at the future instant $n+1$. Therefore, this model considers that the attitudes' reaction function of the social group changes while time goes on, according to some law $L$. This law determines the evolution in the way that the group will react in the future. One can control that evolution in two modes:

• controlling the state $X$, that is, either modifying some or all the agents that cause a reaction of the group (intra or extra individuals, intra or extra social group), or modifying the way in which the social group reacts to those agents, or add more agents, or suppress others, or modify all factors at the same time. This is called in mathematics a *spacial change*, but it is just a jump of the "point" $X$ in the space $\aleph$, without a change of this space itself.

• controlling the dynamical law $L$, that is modifying the deterministic rules in which the attitudes' reacting function will evolve in the future, without changing at the present time, the agents nor the way in which the social group reacts to those agents. This is called in mathematics a *structural change*. It is a change of the rules, according to which the "point" $X$ will evolve in the space $\aleph$, until other structural change is done, and without changing the space $\aleph$ itself, nor the distribution of its points at the present time.

The systems for which *small* spacial changes do not modify the evolution, are called *Lyapunov stable*. The systems for which arbitrarily small spacial changes in some direction modify the evolution, are called *expansive, or Lyapunov unstable, or chaotic.* The systems such that small structural changes do not modify the evolution are called *structurally stable*. Finally, the systems such that arbitrarily small structural changes modify the evolution are called *bifurcating*.

There exist simple mathematical examples that are not chaotic nor bifurcating, other that are not chaotic but  bifurcating, and other, not so simple, that are chaotic but not bifurcating. During many years, there was a conjecture asserting that generic chaotic systems were structurally stable (namely, not bifurcating), until some complicated abstract dynamical systems were invented (Newhouse 1979), which exhibited chaos and a mostly abundant bifurcating behavior, simultaneously.  These are called *wild* systems.

On the contrary of wildeness, chaos is usually structurable stable. In fact, it is a known theorem in mathematics, the following assertion:

Under the hypothesis of uniform hyperbolicity (namely, the existence of uniform exponential rates of expansion of distances among different states of the system), and under the assumption of finite dimension of the space, *the chaotic systems,* such called *Anosov systems* (Anosov 1962) *are structurally stable*. This theorem means that there is an abundant family of mathematical systems such that, even being chaotic, have such a persistent future evolution that would need a relatively *very large* structural change, to behave differently.

Nevertheless, most known theorems about chaotic systems, structural stability and bifurcations, have not still been generalized to infinite-dimensional spaces.  At the same time, most functional spaces, such as the one in our hypothetical example,  have infinite dimensions. Summarizing, to consider abstract structures in mathematics, as for instance abstract functional spaces,  instead of only finite words or matrixes of real numbers, is a very powerful tool which may model very complex systems. It fits to descriptive, qualitative and non quantitative complex systems. Nevertheless, at one hand, the deterministic mathematical hypothesis must be justified to hold in the concrete applied system, and on the other hand, most theorems about deterministic chaos and structural stability, that were proved up to now a days, should be generalized to fit with those applied systems.

## 5.2    Decoding the mathematical results

In paragraph 4.1 we denoted the  deductive relationship of a mathematical proved theorem as

(T)   $A \Rightarrow B$.  If $A$  is  the  mathematical  model  of  a  psychological  system  $\widehat{A}$,  then  the

mathematical thesis $B$ can be decoded into attributes of that system. We denote the decoded result of attributes as $\widetilde{B}$ (see Figure 1.) The decoding or interpretation action from $B$ to $\widetilde{B}$ is writen as:

$$\widetilde{B} \leftarrow B$$

The decoding is the final action in the meta-theory, which leads to the conclusions of the interdisciplinary reasearch that is schematized in Figure 1.

Concrete examples of such decoding process, from thesis of theorems to attributes of psychological systems, can be found in the book of Nowak and Vallacher (1998), and in the articles Ayers (1997) and Robertson (1995). More generally, in the book of Strogatz (1994), very interesting applications to biology, physics, chemistry and engineering are explained with detail. In the two books referred above there is also a review of many mathematical dynamical features of general chaotic and non linear systems, in a context which is directed to a wide audience of scientists of different disciplines.

## 6.    CONCLUSIONS    ▲

Mathematics has still rather few knowledge of chaotic dynamical systems, reduced to dynamics evolving in spaces of relatively low dimensions and that have good regularity properties such as differentiability for instance. Thus, with more reasons, mathematics has not still many proved theorems about most dynamical systems, for instance those that are non linear and evolve in spaces with very large finite or infinite dimension, and those that have discontinuities, like some neuron networks models. Those complex systems appear when modeling some dynamical systems coming from other sciences, in particular from neuroscience and psychology. So, the translation of complex dynamical models from psychology will surely pose new open questions to mathematics. New concepts and mathematical strategies for the proofs of new theorems, should be developed. It is a historical role and motivation for mathematicians, to create and innovate in mathematics, adapting their research agenda to the problems posed from other disciplines. So, we conclude that it is not only mathematical psychology and applied mathematics, which are creating new strategies of research to apply the theory of deterministic chaos, but also pure mathematics, which revises and widens its scope to adapt to the interdisciplinary investigation.

**REFERENCES**

Anosov, D.V. (1962) Structural stability of geodesic flows on compact Riemannian manifolds of negative curvature. *Dokl. Akad. Nauk. SSSR* Vol.145 pp 707—709  ▲

Ayers, S. (1997) The application of chaos theory to psychology. *Theory and Psychology,* Vol. 7 Nº. 3, pp 373—398  ▲

Boole, G. (1854) *An investigation of the laws of thought.* Edition in 1982 translated to Spanish *Investigación sobre las leyes del pensamiento.* Ediciones Paraninfo. Madrid.  ▲

Budelli, R; Catsigeras, E., Enrich, H., Torres, J. (1991) Two neuron network. *Biological Cybernetics*, v. 66, p. 95-101, 1991.  ▲

Budelli, R; Catsigeras, E (1992) Limit Cycles in a model of bineuronal networks. *Physica D Nonlinear Phenomena*, Vol. 56, p. 235-252, 1992.  ▲

Budelli, R; Catsigeras, E; Rovella, A; Gómez, L. (1997): Dynamical behavior of pacemaker neurons networks. *Journal of Nonlinear Analysis*, Vol. 30. Nº 3, pp. 1633—1638.  ▲

Camacho, L. (2006) La lógica en Kant y en George Boole. {The logic in Kant and in George Boole} *Rev. Filosofía Univ. Costa Rica* 111-112, pp 49—56  ▲

Catsigeras, E. (2010): Chaos and stability in a model of inhibitory neuronal network. . *International Journal of Bifurcation and Chaos*, Vol. 20 Nº 2 , pp. 349—360.  ▲

Cessac, B. (2008) A discrete time neural network model with spiking. *Journal of Mathematical Biology* Vol. 54, pp 311—345  ▲

Coombes, S. and Lord, G.J. (1997) Desynchronization of pulse-coupled integrate-and-fire neurons. *Physical Review E*, Vol. 55, Nº 3.  ▲

Coombes,S. (2007). Mathematical neuroscience. *Journal of Mathematical Biology*, Vol.54,  pp. 305—307.  ▲

Coombes,S. and C R Laing,C.R. (2009). Delays in activity based neural networks. *Philosophical Transactions of the Royal Society A*, Vol. 367, pp 1117—1129.  ▲

Cooper,L.N. (1995): *How we learn. How we remember. Toward an understanding of brain and neural systems.* World Scientific. Singapur.  ▲

Eliasmith, C. (1996) The third contender: A critical examination of the Dynamicist Theory of Cognition *Philosophical Psychology*, Vol. 9, Issue 4 pp. 441—463  ▲

Feudel, U., Neiman, A., Pei,X., Wojtenek,W., Braun, H., Huber, M., Moss, F.: Homoclinic bifurcation in a Hodgkin-Huxley model of thermally sensitive neurons. *Chaos. An Interdisciplinary Journal of Nonlinear Science.* Vol. 10, 231 (2000) doi:10.1063/1.166488   ▲

Goldstein, J (1995) The Tower of Babel in Nonlinear Dynamics: toward the Clarification of Terms. *Chaos Theory in Psychology and the Life Sciences*, Robertson,R. and  Combs, A. (editors) Lawrence Erlbaum Assoc.Inc. Publ. New Jersey. pp. 39-48  ▲

Hodgkin, A.L.and Huxley, A.(1952): A quantitative description of membrane current and its application to conduction and excitation in nerve. *Journal of Physiology,* Vol. 117, nº 4, pp. 500—544. ▲

Izhikevich, E. (2007) *Dynamical systems in neuroscience: the geometry of excitability and bursting.* MIT Massachusetts Institute of Technology Press. Cambridge. ▲

Kellert, S.H (2008) *Borrowed Knowledge: Chaos Theory and the Challenge of Learning across Disciplines.* The University of Chicago Press, Chicago and London ▲

Kohonen,T. (1977): *Associative Memory: A system-theoretical approach.* Springer. Berlin. ▲

Lamberti, P, Rodríguez, V. (2007) Desarrollo del modelo matemático de Hodgkin y Huxley en neurociencias. {Develpment of the mathematical model of Hodgkin and Huxley in neurosciences} *Revista Electroneurobiología,* Vol. 15 Nº 4. pp 31—60
http://electroneubio.secyt.gov.ar/Lamberti-Rodriguez_Hodgkin-Huxley.htm (21/06/2010) ▲

Lansner, A. (2009) : Associative memory models: from the cell-assembly theory to biophysically detailed cortex simulations. *Trends in Neurosciences.* Vol. 32. Nº3. pp 178—186. ▲

Lewowicz, J. (1990) Expansive Homeomorphisms of Surfaces. Boletim da Sociedade Brasileira de Matemática. Vol. 20 Nº. 1 pp 113 —133. ▲

Lewowicz, J. (2002) Acerca del Caos. {About Chaos} *Actas de Fisiología* Vol. 8, pp 41—53. ▲

Lewowicz, J. (2008) Caos determinista y expansividad. {Deterministic Chaos and Expansivity} *Anales de la Academia de Ciencias Exactas, Físicas y Naturales Argentina.* ▲

Lorenz, E. (1995) *La esencia del caos. {The essence of chaos}* Editorial Debate. Madrid. ▲

Mañé. Ricardo (1983) *Ergodic Theory and Differentiable Dynamics* (1986). Original edition in Portuguese (1983): *Teoría Ergódica.* IMPA, Projeto Euclides, Rio de Janeiro. ▲

Markarian, R. & Gambini, R. (editors) (1997) *Certidumbres. Incertidumbres. Caos. Reflexiones en torno a la ciencia contemporánea.{Certainties. Uncertainties. Chaos. Reflexions around the contemporary science.}* Ediciones Trilce. Montevideo. ▲

Massera, J.L. (1997) Reflexiones de un matemático sobre la dialéctica. {Reflexions of a mathematician about the dialectic} *Pre-publicaciones de Matemática de la Universidad de la República, Uruguay.* Nº 97/01 ▲

Massera, J.L. (1988) Problemas de filosofía de la matemática, de sus fundamentos y metodología. {Philosophical problems of the mathematics, its foundations and its methodology. } *Publicaciones Matemáticas del Uruguay.* Vol. 1, pp 11—26. ▲

Mirollo, R. And Strogatz, S. (1990): Synchronization of pulse-coupled biological oscillators. *SIAM Journal of Applied Mathematics.* Vol. 50 Nº. 6. pp 1645-1662. ▲

Mizraji,E. (2007): Redes Neuronales. (Neuronal Networks.} In F. Simini (editor) *Ingeniería Biomédica {Biomedic Engineering}.* Publicaciones de la Universidad de la República. Montevideo. ▲

Mizraji, E.(2008) Neural Memories and Search Engines. *International Journal of General Systems* Vol. 37 Nº.6 , pp. 715—738. ▲

Mizraji, E. (2010) *En busca de las leyes del pensamiento. {In the search of the thinking laws.}* Ediciones Trilce. Montevideo. ▲

Newhouse, S. (1979) The abundance of wild hyperbolic sets and non-smooth stable sets for diffeomorphisms. *Publications Mathématiques de L'IHÉS* Vol. 50, N° 1 pp 101-151  ▲

Nowak, A. and Vallacher, R. (1998) *Dynamical Social Psychology.* The Guilford Press, New York. ▲

Rieke,F., Warland, D. , de Ruyter, R., Steveniorck, V. and Bialek, W. (1997): *Spikes. Exploring the neural code.* MIT Massachusetts Institute of Technology Press. Cambridge. ▲

Robertson, R. (1995): Chaos Theory and the Relationship Between Psychology and Science. *Chaos Theory in Psychology and the Life Sciences*, Robertson,R. and Combs, A.(editors) Lawrence Erlbaum Assoc. Inc. Publ. New Jersey, pp. 3—16. ▲

Ruelle,D. (1990): Deterministic chaos: the science and the fiction. *Proceedings  of the Royal Society of London  A* Vol. 427 pp 241—248  ▲

Ruelle, D. (1993). *Azar y caos.{Chaos and hazard.}* Editorial Alianza. Madrid. ▲

Rummens,S., Cuypers,S. (2010): Determinism and the paradox of predictability. *Erkenntnis* Vol. 72 pp 233—249 ▲

Scott, B. (1994) Chaos, self-organization, and psychology. *American Psychologist*, Vol 49(1), pp. 5-14. ▲

Searle, J. (1978) Literal Meaning *Erkenntnis* 13 pp. 207—224.  ▲

Sokal A., Bricmont, J. (1999): *Imposturas intelectuales. {Intelectual impostures}* Ed. Paidós. Barcelona.  ▲

Stewart, I. (1989) *Does God play dice? The new mathematics of chaos.* Ed. Basil Blackwell. London. ▲

Strogatz, S.H. (1994) *Non Linear Dynamics and Chaos. With Applications to Physics, Biology, Chemistry, and Engineering.* Perseus Publshing, Cambridge  ▲

Timme, M., Wolf, F. and Geisel T. (2002): Coexistence of regular and irregular dynamics in complex networks of pulse-coupled oscillators. *Physical Review Letters*. Vol. 89, N° 25.  ▲

Vallacher, R. and Nowak, A. (1997) The Emergence of Dynamical Social Psychology *Psychological Inquiry* Vol. 8,  N°.2, pp 73—99. ▲

Vieitez, José L. (1996) Expansive homeomorphisms and hyperbolic diffeomorphisms on 3-manifolds. *Ergodic Theory and Dynamical Systems* Vol. 16 N°.3  pp 591—622 ▲

Von Neumann,J. (1958-posthumous): *The computer and the brain.* 2nd. Edition (2000): *The computer and the brain.  With a foreword* by P.M. Churchland and P.S.Churchland. Yale University Press. New Haven and London. ▲

Yang, T. and Chua, L. (1997) Impulsive stabilization for control and synchronization of chaotic systems: theory and application to secure communication. *IEEE Transactions on circuits and systems E*, Vol. 44, N°. 10.  ▲

Zollman, K. (2010) The epistemic benefit of transient diversity. *Erkenntnis* Vol 72 pp. 17-35  ▲